\documentclass[aps,prc,twocolumn,floatfix,superscriptaddress,nofootinbib,preprintnumbers,longbibliography,amsmath,amsthm,amssymb]{revtex4-2}
\usepackage{graphicx}

\usepackage{amsfonts}
\usepackage{color}
\usepackage[colorlinks=true,linkcolor=blue,citecolor=blue]{hyperref}
\usepackage{dcolumn}% Align table columns on decimal point
\usepackage{bm}% bold math
\usepackage{braket}

\begin{document}
\allowdisplaybreaks[1]
\title{Nuclear mass table in density functional approach inspired by neutron-star observations}
\author{Hana Gil}
\affiliation{Center for Extreme Nuclear Matters, Korea University, Seoul 02841, South Korea}
\author{Nobuo Hinohara}
\affiliation{Center for Computational Sciences, University of Tsukuba, Tsukuba, Ibaraki 305-8577, Japan}
\affiliation{Faculty of Pure and Applied Sciences, University of Tsukuba, Tsukuba, Ibaraki 305-8571, Japan}
\author{Chang Ho Hyun}
\affiliation{Department of Physics Education, Daegu University, Gyeonsan 38453, South Korea}
\author{Kenichi Yoshida}
\email[E-mail: ]{kyoshida@rcnp.osaka-u.ac.jp}
\affiliation{Research Center for Nuclear Physics, Osaka University, Ibaraki, Osaka 567-0047, Japan}
\affiliation{RIKEN Nishina Center for Accelerator-Based Science, Wako, Saitama 351-0198, Japan}
\date{today}% It is always \today, today,
\begin{abstract}
\begin{description}
\item[Background]
Nuclear energy-density functional (EDF) approach 
has been widely used to describe nuclear-matter equations of state (EoS) and properties of finite nuclei. 
Recent advancements in neutron-star (NS) observations 
have put constraints on the nuclear EoS. 
The Korea--IBS--Daegu--SKKU (KIDS) functional has been then developed to satisfy the NS observations and 
applied to homogeneous nuclear matter and spherical nuclei. 
\item[Purpose]
We examine the performance of the KIDS functional 
by calculating the masses and charge radii 
of even-even nuclei towards the drip lines.
\item[Method] 
The Kohn--Sham--Bogoliubov equation is solved by taking into 
account the axial deformation.
\item[Results]
The root-mean-square deviation of the binding energy 
and the charge radius for the KIDS functional is 
$4.5$--$5.1$ MeV and $0.03$--$0.04$ fm, 
which is comparable to that for existing EDFs.
The emergence and development of nuclear deformation in open-shell nuclei are well described. 
The location of the neutron drip line is according to the nuclear-matter parameter characterizing the low-mass NS.
\item[Conclusions]
The NS-observation-inspired EDF offers a reasonable reproduction of the structures of finite nuclei. 
A future global optimization including more nuclear data will 
give better accuracy and high predictive power of neutron-rich nuclei.
\end{description}
\end{abstract}
\maketitle
\section{Introduction}\label{intro}
A microscopic construction of 
the nuclear equation of state (EoS) 
and its determination from nuclear experiments
have been an interdisciplinary issue between nuclear physics and astrophysics~\cite{Roca-Maza:2018ujj,Lattimer_2013,RevModPhys.89.015007}.
Thanks to the advance in astronomical observations, data on the properties of a neutron star (NS) have become more rich, diverse, and precise~\cite{Demorest2010,Steiner_2010,Steiner_2013,doi:10.1126/science.1233232,Cromartie2019,Riley_2019,Miller_2019}.
One can see some attempts to determine the EoS directly 
from the observations by using machine learning~\cite{Farrell:2022lfd,PhysRevD.101.054016,Fujimoto2021,Morawski:2020izm}. 
Bridging such recent data and nuclear structure properties 
is a new challenge in nuclear physics.
A desired model, which can describe both infinite nuclear matter and finite nuclei, 
must be able not only to reproduce the existing data accurately but also to be systematic 
at improving its predictive power and flexible to the addition of residual forces.

As an intermediate step, 
the KIDS (Korea-IBS-Daegu-SKKU) density functional was applied to constrain 
the EoS 
by using state-of-the-art data on X-ray sources, low-mass X-ray binaries, and gravitational waves~\cite{prc2021, npsm2021, ijmpe2022}.
The symmetry energy thus determined is consistent with the results in the literature, and the ranges of the uncertainty could be reduced non-negligibly.
In subsequent work, an effect of the symmetry energy has been explored in finite nuclei by considering the Nd isotopes
in the neutron-rich region~\cite{nd2022}.
Used in the work were four models, KIDS-A, B, C, and D, which have distinctive stiffness of the symmetry energy.
The models agree well with the data on binding energy, charge radius, and quadrupole deformation.
On the other hand, predictions in the neutron-rich regions such as the neutron skin thickness and the neutron drip line show a strong dependence on the symmetry energy.
For a better understanding of the behavior of the model and the effect of the symmetry energy,
extension of the analysis to the whole nuclear landscape is in due order.

In the present work, 
we investigate the properties of even-even nuclei for the atomic number $Z$ from 8 to 110,
and for the neutron number $N$ from the proton drip line to $N=3Z$.
Major concerns are 
1) How well the KIDS model thus constructed by using the NS data works across the entire range of the nuclear chart,
and 2) To explore the dependence on 
the symmetry energy and the ranges of its uncertainty of 
the KIDS model constrained by the modern NS data. 
We analyze the results for the binding energy, charge radius, quadrupole deformation, and neutron drip line.
The mean accuracy of the model is similar to that of existing Skyrme models, but it is obviously inferior to the globally fitted mass models.
The result opens a challenge to the KIDS density functional whether it can achieve an accuracy comparable to the
globally fitted mass models.

The paper is organized in the following order.
In Sec.~\ref{model}, we briefly introduce the models.
Some details about the numerical calculations are also summarized.
In Sec.~\ref{results}, we present the results and discuss them.
We summarize the work in Sec.~\ref{summary}.

\section{Model}\label{model}

%Table 1
\begin{table}[t]
\caption{\label{tab:model} Parameters of the nuclear matter EoS 
including the compression modulus $K_0$, and the symmetry energy parameters $J$, $L$, and $K_\tau$ are given in the units of MeV.}
\begin{ruledtabular}
\begin{tabular}{ldddd}
 & K_0 & J & L & K_\tau \\ \hline
KIDS-A & 230 & 33 & 66 & -420 \\
KIDS-B & 240 & 32 & 58 & -420 \\
KIDS-C & 250 & 31 & 58 & -360 \\
KIDS-D & 260 & 30 & 47 & -360 \\
SLy4    & 229.9 & 32.0 & 45.9 & -322.8 
\end{tabular}
\end{ruledtabular}
\end{table}

In this work, we use four KIDS functionals, 
KIDS-A, KIDS-B, KIDS-C, and KIDS-D.
There are nine parameters in the functional; 
seven parameters of 
the functional are adjusted to three nuclear saturation properties (the saturation density $\rho_0=0.16\, {\rm fm}^{-3}$, the binding energy per nucleon 16 MeV, and the incompressibility $K_0=230\text{--}260$ MeV) and
$R_{1.4} = 11.8\text{--}12.5$~km where $R_{1.4}$ denotes the radius of a NS with the mass $1.4 M_\odot$; see Table~\ref{tab:model} for the 
nuclear-matter parameters.
The remaining two are fitted to six nuclear data (the energy per particle and the charge radius of $^{40}$Ca, $^{48}$Ca, and $^{208}$Pb).
Two more parameters are added in the pairing functional for neutrons and protons.
The pairing parameters are adjusted to the three-point formula for the odd-even staggering centered at $^{156}$Dy.
The total number of parameters in each model is 11, and their numerical values can be found in Ref.~\cite{nd2022}.

The KIDS models are implemented in the {\sc hfbtho} code~\cite{sto13} to solve 
the Kohn--Sham--Bogoliubov equation taking the axial deformation into account. 
The calculations are performed in the $N_{\text{max}}=20$ full spherical oscillator shells. 
To find the global-minimum solution, we start calculations 
from the initial configurations with the mass quadrupole deformation $\beta_2 = 0$ and $\pm 0.3$.
The quasiparticle (qp) states are truncated according to the equivalent single-particle energy cutoff at 60 MeV.

The KIDS functionals have been applied to 
the quasielastic scattering of an electron and a neutrino off a nucleus~\cite{prc104, prc105, plb833}.
It is shown that the cross-section is sensitive to and depends critically on the in-medium effective mass of the nucleon.
The result demonstrates that the lepton-nucleus reactions can provide a unique channel to probe
the nuclear dynamics and nuclear matter properties.
The model gives the neutron skin thickness of $^{48}$Ca and $^{208}$Pb consistent with the measurement from 
the dipole polarizability and interaction cross section~\cite{nskin2022}.

\section{Results and discussion}\label{results}

\subsection{Binding energy}

%% Figure 1 %%
\begin{figure*}
\begin{center}
\includegraphics[width=8.5cm]{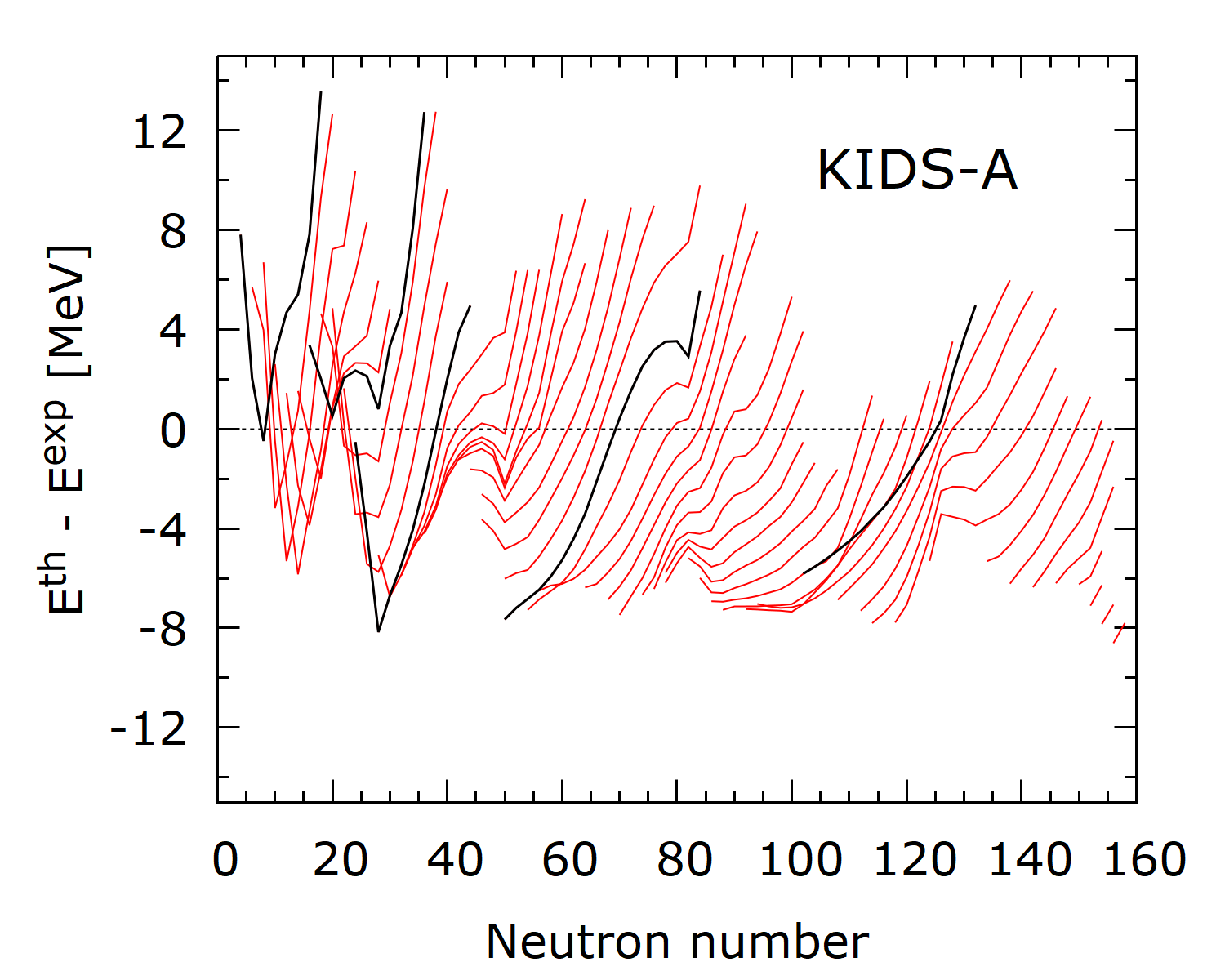}
\includegraphics[width=8.5cm]{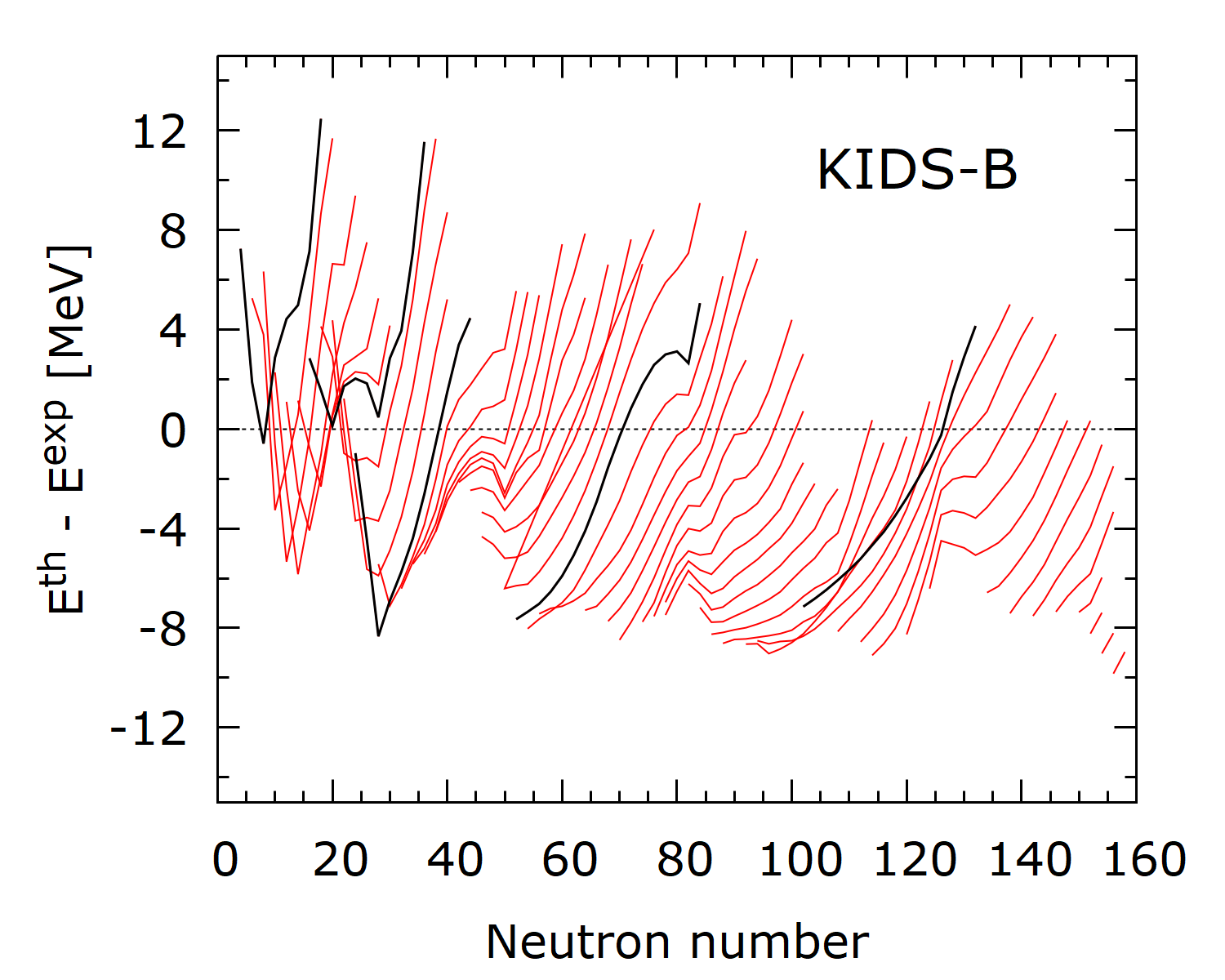}
\includegraphics[width=8.5cm]{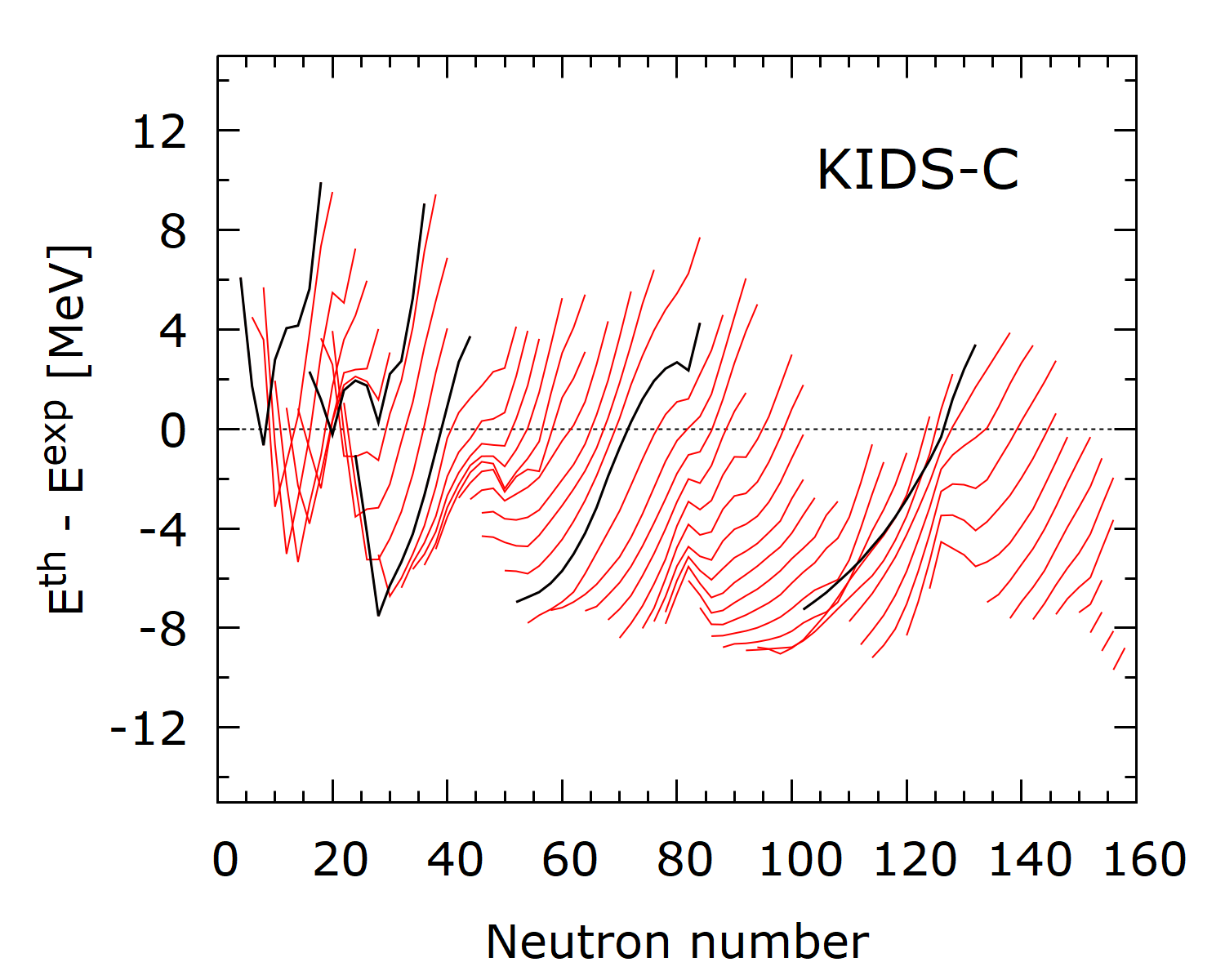}
\includegraphics[width=8.5cm]{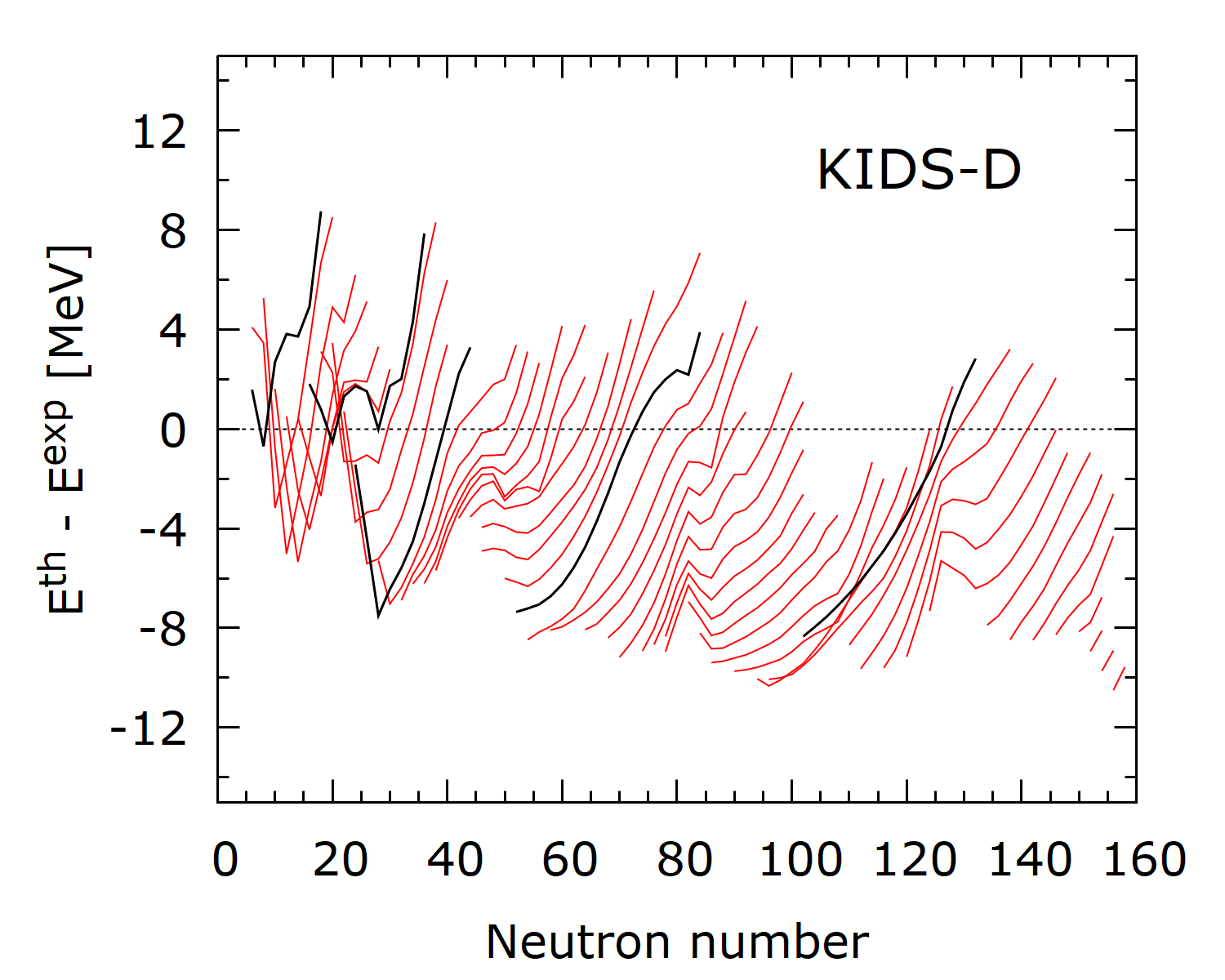}
\end{center}
\caption{
\label{fig:mdiff}
Binding energy residuals between the KIDS results and experiment for 632, 630, 628, and 623 even-even nuclei 
for the KIDS-A, KIDS-B, KIDS-C, and KIDS-D models, respectively. 
The residuals are calculated only for the bound nuclei in each model among the available experimental data.
Black lines denote the results for O, Ca, Ni, Sn, and Pb isotopes, respectively.
}
\end{figure*}

Figure~\ref{fig:mdiff} shows the calculated binding energies subtracted by the experimental data~\cite{ame2020}.
Here, the binding energy in Fig.~\ref{fig:mdiff} is given by
\begin{equation}
{\rm BE}(N,Z) = NM_n+ZM_p-M(N,Z),
\end{equation}
where $M_n$ and $M_p$ are neutron and proton masses, and $M(N,Z)$ is the nuclear mass.
The models KIDS-A--D show very similar patterns.
Black lines denote the isotopic chains of O, Ca, Ni, Sn, and Pb.
Since the models are fitted to the binding energy of $^{40}$Ca, $^{48}$Ca, and $^{208}$Pb,
two downward spikes in the line of Ca and $N=126$ in the line of Pb are very close to zero.
It is notable that decreasing and increasing behaviors are mixed in an isotopic chain for $N \lesssim 30$,
but for $N\gtrsim30$ each line shows a simply-increasing behavior. 
In the light $N=Z$ nuclei between Ne and Ar, 
the calculation underestimates the binding energy, 
which results in the cusp. 
In heavier nuclei with $N>30$, the cusp appears at the magic numbers.
Slopes of the isotopic lines for a given nuclide look similar among the models KIDS-A--D.
In Ref.~\cite{nd2022}, we investigated the origin of the increasing behavior of the residual as the neutron number increases
in the isotopic chain of Nd.
We found that by controlling the slope parameter $L$ of the symmetry energy,
it is possible to make the line of the Nd isotopes either flat or stiff.
We thus expect that we can construct a mass model based on a KIDS-EDF with appropriate parameters.

We estimate the accuracy of the model in two ways.
One is to evaluate the standard root-mean-square deviation (RMSD) defined by
\begin{equation}
{\rm RMSD}(O) = \sqrt{\left< (O_{\rm th}  - O_{\rm exp})^2 \right>},
\end{equation}
where $O$ denotes an observable.
Another measure is a comparison in the logarithmic scale defined by
\begin{equation}
R(O) = 100 \times \ln \left(\frac{O_{\rm th}}{O_{\rm exp}} \right).
\end{equation}

%%%%% Table2 %%%%%
\begin{table}[t]
\begin{center}
\caption{\label{tab1}
RMSD and $R$ value of the KIDS models.
RMSD($E$)
and RMSD($R_c$) are in the unit of MeV
and fm, and 
$\left<R(E)\right>$ and $\left<R(R_c)\right>$ are dimensionless.}
\begin{ruledtabular}
\begin{tabular}{ccccc}
Model & KIDS-A & KIDS-B & KIDS-C & KIDS-D   \\ \hline
RMSD($E$) & 4.50 & 4.86 & 4.70 & 5.08  \\
$\left<R(E) \right>$ & 0.011 & $-$0.073 & $-$0.127 & $-$0.211  \\ 
RMSD($R_c$) & 0.0323 & 0.0356  & 0.0345 & 0.0384  \\
$\left<R(R_c)\right>$ & $-$0.232 & $-$0.411 & $-$0.457 & $-$0.602 
\end{tabular}
\end{ruledtabular}
\end{center}
\end{table} 

Table~\ref{tab1} summarizes the RMSD and $R$ values of the KIDS model.
The number of data included in the evaluation is 
632, 630, 628, and 623 for the KIDS-A, 
KIDS-B, KIDS-C, and KIDS-D models, respectively.
RMSD($E$) values of the KIDS models are in the range $4.5$--$5.1$ MeV.
The accuracy is similar to that of the SLy4 model (4.8 MeV),
but larger than globally fitted mass models, e.g., 1.45 MeV of UNEDF0~\cite{PhysRevC.82.024313}.
We evaluate the average of $R$ values over the results.
Mean values of $R(E)$ in Table~\ref{tab1} indicate that 
the calculated binding energy deviates
from the experimental data by 
0.011\%, $-0.073$\%, $-0.127$\% and $-0.211$\%
on the average for the KIDS-A, KIDS-B, KIDS-C, and KIDS-D models, respectively.

\subsection{Charge radius}

%% Figure 2 %%
\begin{figure*}
\begin{center}
\begin{tabular}{cc}
\includegraphics[width=9cm]{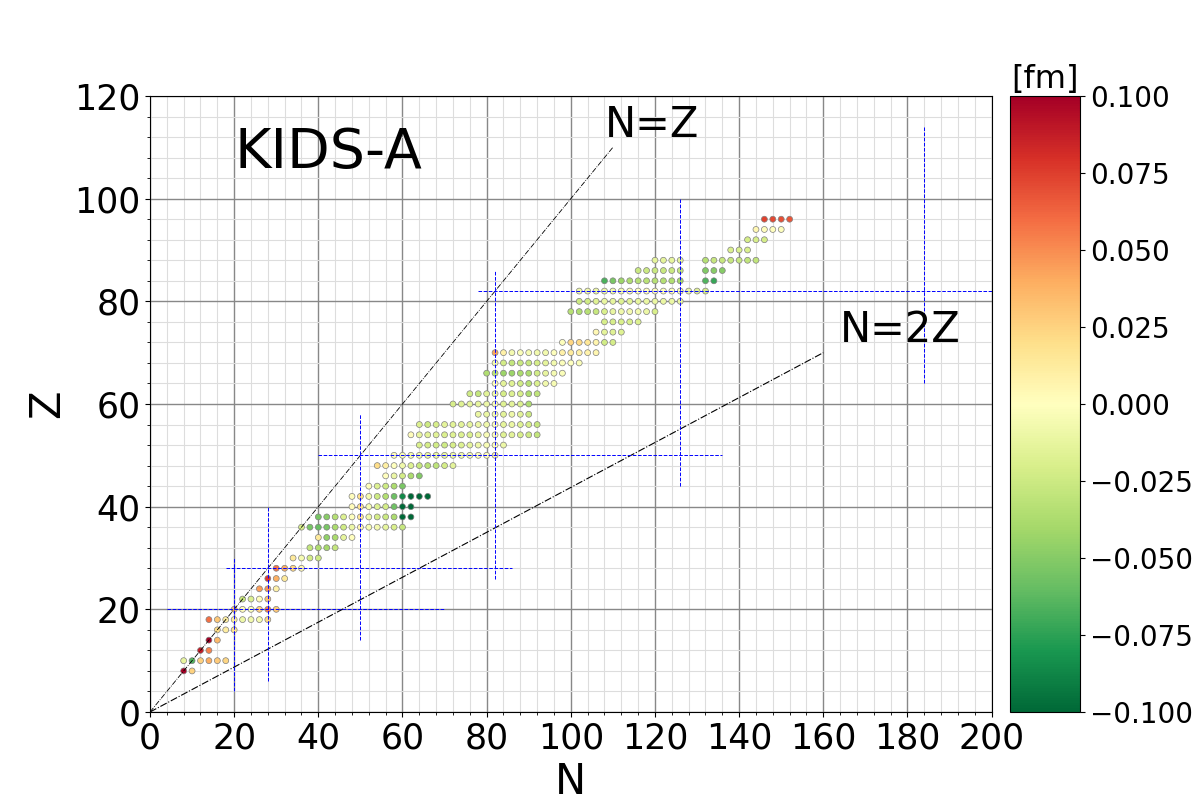} &
\includegraphics[width=9cm]{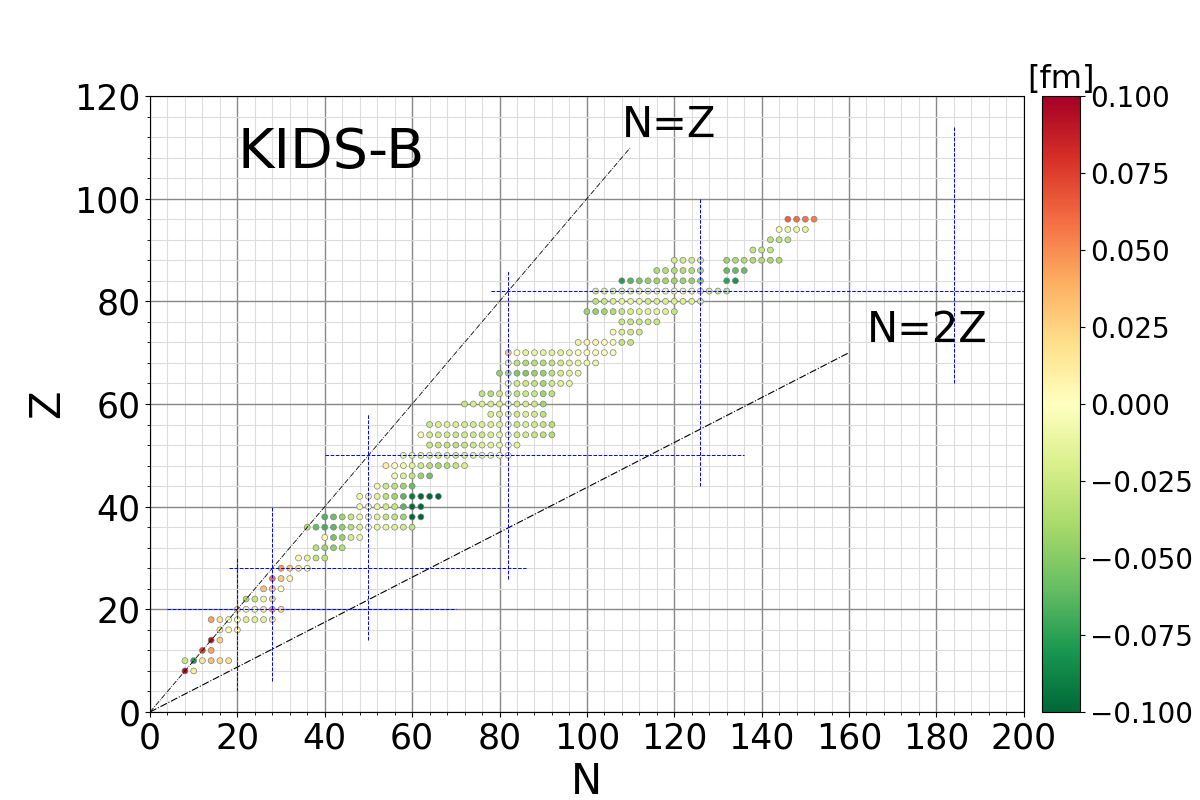} \\
\includegraphics[width=9cm]{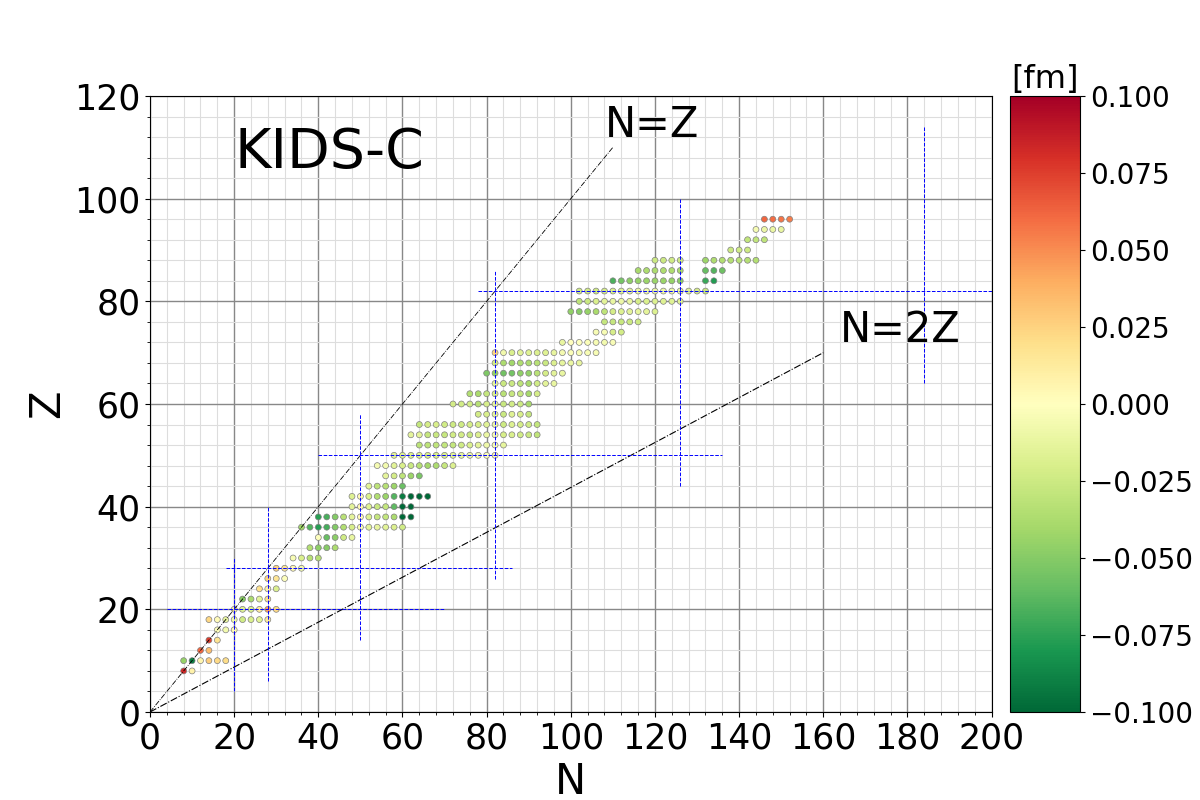} & 
\includegraphics[width=9cm]{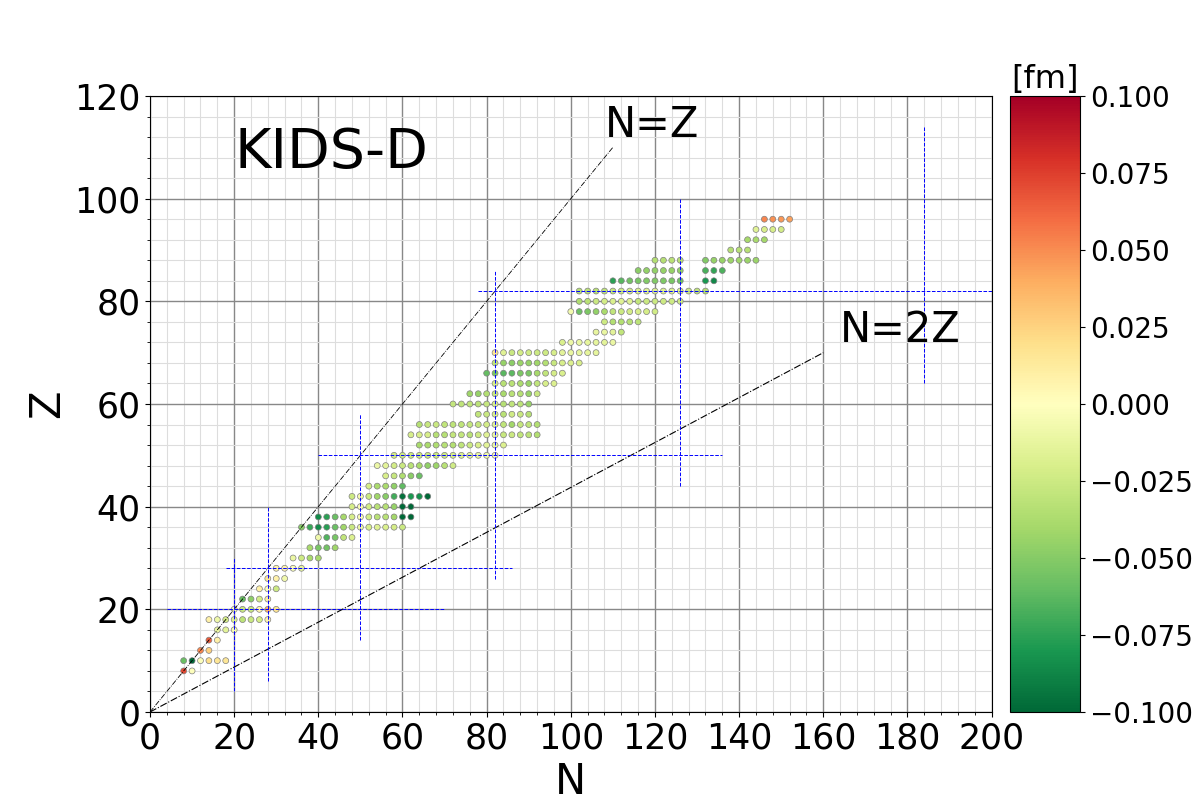} 
\end{tabular}
\end{center}
\caption{Calculated charge radius difference from 
the experimental value for 344, 344, 343, and 343 even-even nuclei 
for the KIDS-A, KIDS-B, KIDS-C, and KIDS-D models, respectively.
}
\label{fig:rcdiff}
\end{figure*}

Figure~\ref{fig:rcdiff} shows the difference 
of charge radii 
$r_{\rm ch}^{\rm th}-r_{\rm ch}^{\rm exp}$. 
The charge radius is evaluated as
\begin{eqnarray}
r_{\rm ch} &=& \sqrt{r^2_p + 0.64 \,\,{\rm fm^2}},
\end{eqnarray}
where $r^2_p$ is the 
expectation value on the HFB vacuum of a point proton radius,
and the contribution from the finite proton size is included \cite{sto13}.
Table~\ref{tab1} summarizes the RMSD and $R$ values for the charge radius.
The number of data included in the evaluation is 
344, 344, 343, and 343 for the KIDS-A, 
KIDS-B, KIDS-C, and KIDS-D models, respectively.
The experimental data are given in Refs.~\cite{ANGELI201369,LI2021101440}.
The deviation 0.032--0.038 fm is 
comparable to 0.032 fm obtained with 
the DRHBc calculation \cite{drhbc2022}.

We find notable deviations in several regions.
Except for light $N \sim Z$ nuclei, 
one sees an appreciable discrepancy in the $Z \sim 40, 84$, and $96$ isotopes.
The charge radius of the Sr, Zr, and Mo isotopes around $N=60$ 
is calculated to be smaller than the measured one. 
This is because the calculations produce a 
spherical or weakly-deformed configuration for the ground state, while a largely-deformed configuration
is suggested experimentally~\cite{PhysRevLett.106.202501,PhysRevC.101.044311}.

The calculated charge radius of the Po isotopes with $N=108$ and $110$ 
is also smaller than the measured value. 
The calculation produces a tiny oblate deformation 
with $|\beta_2| < 0.1$
whereas the SLy4 model predicts 
the oblate deformation with $|\beta_2|\sim 0.2$~\cite{PhysRevC.68.054312}. 

On the other hand, the calculated charge radius 
of the Cm isotopes with $N=146\textrm{--}152$ is larger than 
the measured one. The deviation is similar to the calculation 
with the PC-PK1 model~\cite{drhbc2022}.

\subsection{Deformation}

%% Figure 3 %%
\begin{figure*}
\begin{center}
\begin{tabular}{cc}
\includegraphics[width=9cm]{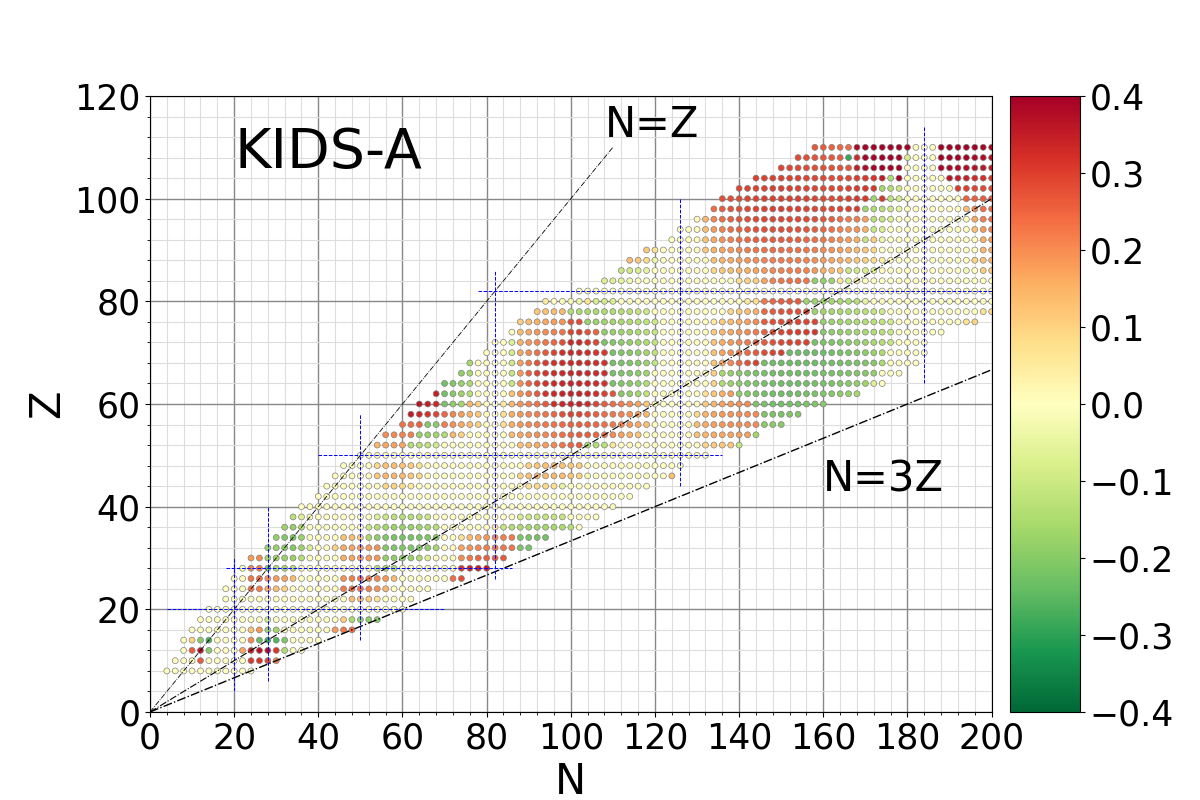} & 
\includegraphics[width=9cm]{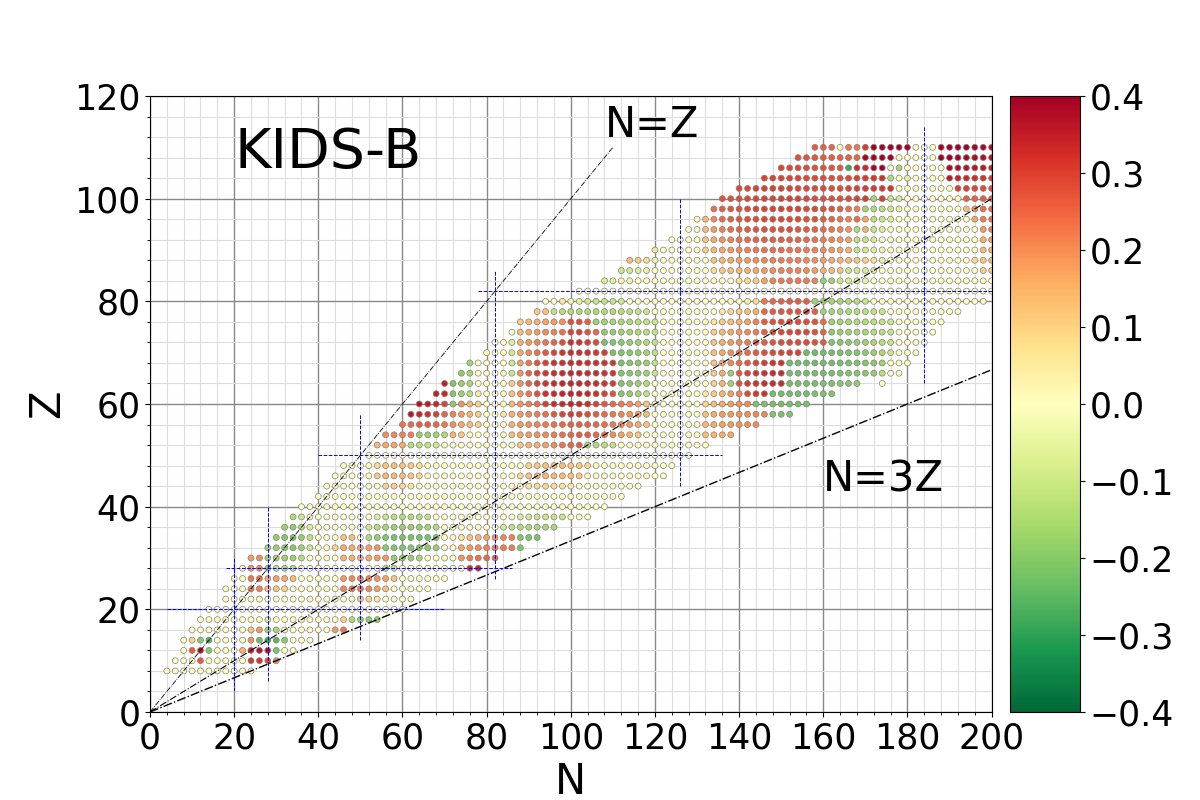} \\
\includegraphics[width=9cm]{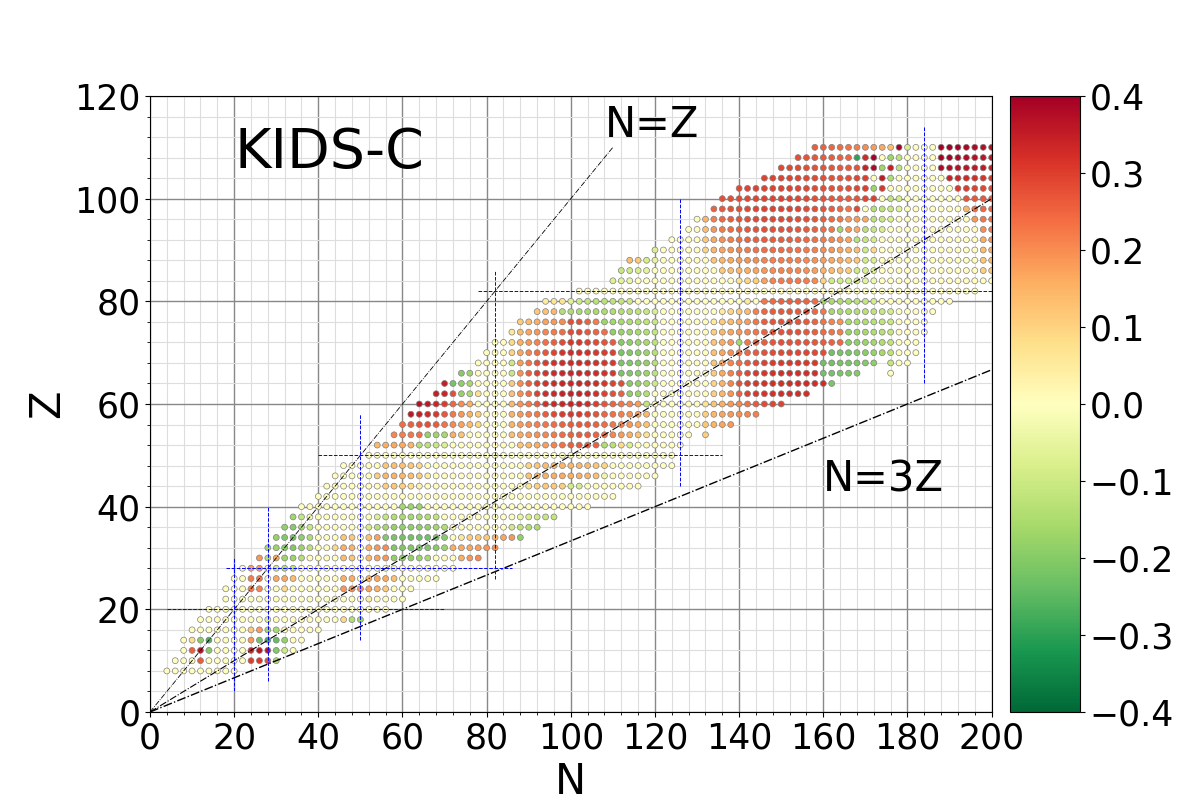} & 
\includegraphics[width=9cm]{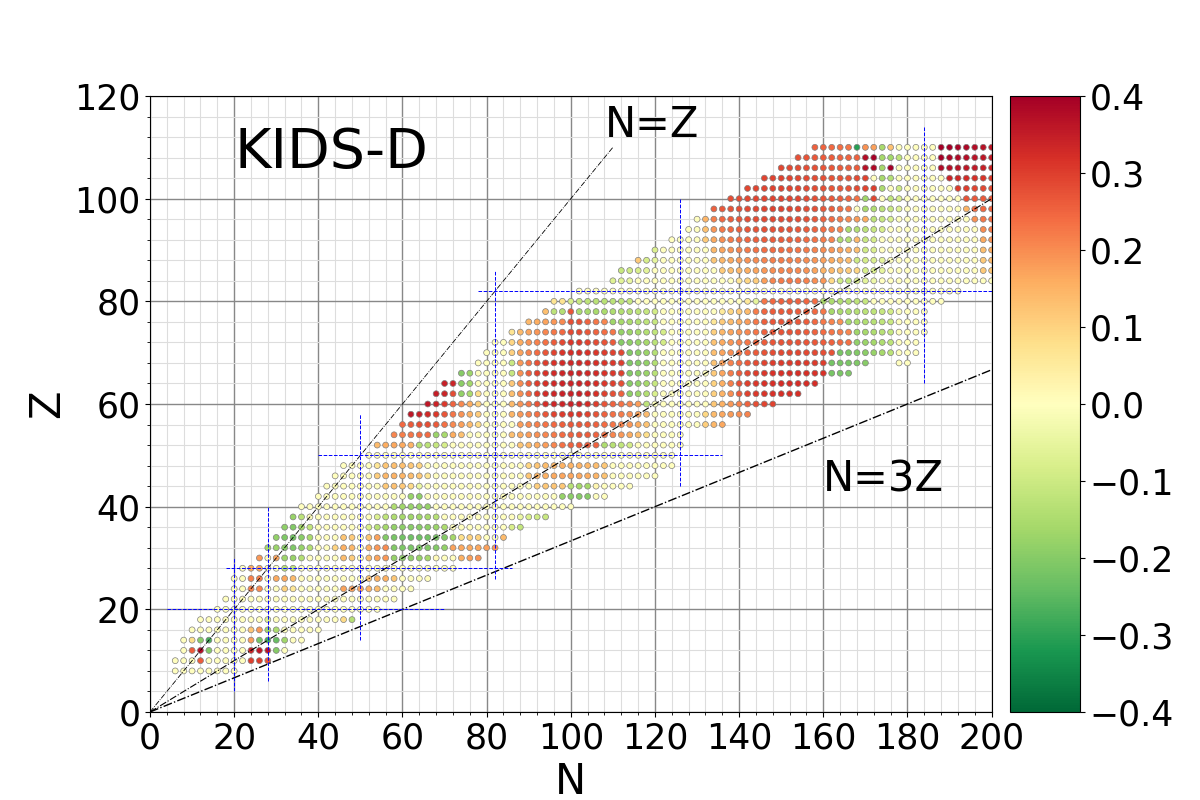}
\end{tabular}
\end{center}
\caption{Calculated quadrupole deformation $\beta_{2,p}$ 
for bound nuclei obtained by employing the KIDS-A--D models.
}
\label{fig:beta}
\end{figure*}

Stepping away from the magic numbers, the deformation appears and then develops by increasing the neutron or proton number. 
Existing EDF calculations have described well 
the evolution of deformation~
\cite{drhbc2022,PhysRevC.81.014303,PhysRevC.68.054312}.
The present KIDS functionals, 
where deformed nuclei are not included in 
determining the coupling constants, 
also describe the onset and 
the development of deformation including 
the sign of quadrupole deformation,
as shown in Fig.~\ref{fig:beta}.
The deformation parameter here is defined as
\begin{align}
    \beta_{2,p} = \sqrt{\frac{\pi}{5}} \frac{ Q_{2,p}}{r_p^2},
\end{align}
where
$Q_{2,p}$ is 
the quadrupole moment of protons.
A gradual change of deformation is produced  
not only in the Nd isotopes, which we have already 
discussed in our previous study~\cite{nd2022}, 
but in other lanthanides.

The disappearance of the magic numbers and 
the appearance of new magic numbers 
in exotic nuclei have been 
discussed both experimentally and theoretically ~\cite{RevModPhys.92.015002}.
The KIDS models give a spherical configuration for the neutron-rich nuclei with $N=20$ 
as other EDF models do~\cite{drhbc2022,PhysRevC.81.014303,PhysRevC.68.054312} 
in contrast to the measurements. 
For $N=28$, the KIDS models describe well the evolution of shape from the oblate to prolate deformations toward a neutron drip line as in Refs.~\cite{PhysRevC.86.051301,Suzuki:2022ngr}.
    In a neutron-deficient side, 
    an oblate configuration appears at $N=28$, and 
    the KIDS-A and B models produce the oblate deformation in $^{56}$Ni.
    
    A possible deformation of the Sn isotopes in the very neutron-rich region around $N=100$ is predicted by some calculations~\cite{PhysRevC.68.054312,PhysRevC.81.014303}. 
    However, the present calculation using the KIDS models predicts 
    the spherical shape for all the Sn isotopes.

\begin{figure}
\includegraphics[width=80mm]{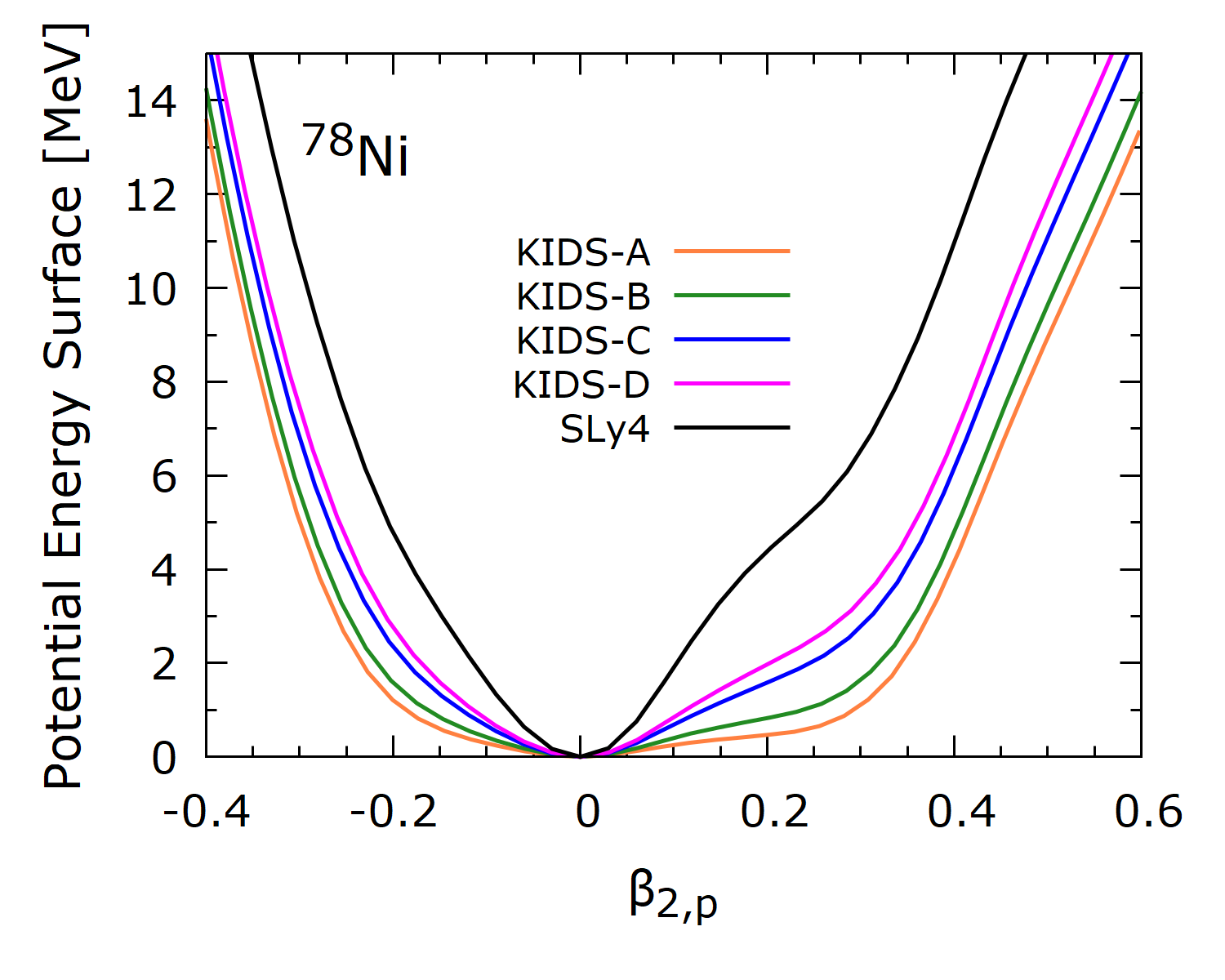}
\caption{
\label{fig:PES_Ni}
Calculated PES of
$^{78}$Ni using the KIDS-A--D models.
The result obtained by using the SLy4 functional 
is included.
}
\end{figure}

    KIDS-A--D models predict the breaking of the $N=50$ and 82 spherical magic numbers near the drip line, 
    while the SLy4 gives the spherical configuration~\cite{PhysRevC.68.054312}. 
    An oblate configuration appears in the ground state at $N\sim 50$, 
    and a prolate configuration shows up at $N\sim 82$.
    The magnitude of deformation is the largest,
    and the deformed region is wide in the KIDS-A model. 
    As an example, we show the 
    potential energy surface (PES)
    of a doubly-magic nucleus $^{78}$Ni in Fig.~\ref{fig:PES_Ni}.
    The ground state is soft against the quadrupole deformation with the KIDS models 
    compared with the SLy4 functional.
    The KIDS-A model gives the softest PES. 

%%% Figure 4 %%%    
%%%%%

\subsection{Drip line}

%%% Figure 5 %%%
\begin{figure}
\begin{tabular}{cc}
\includegraphics[width=90mm]{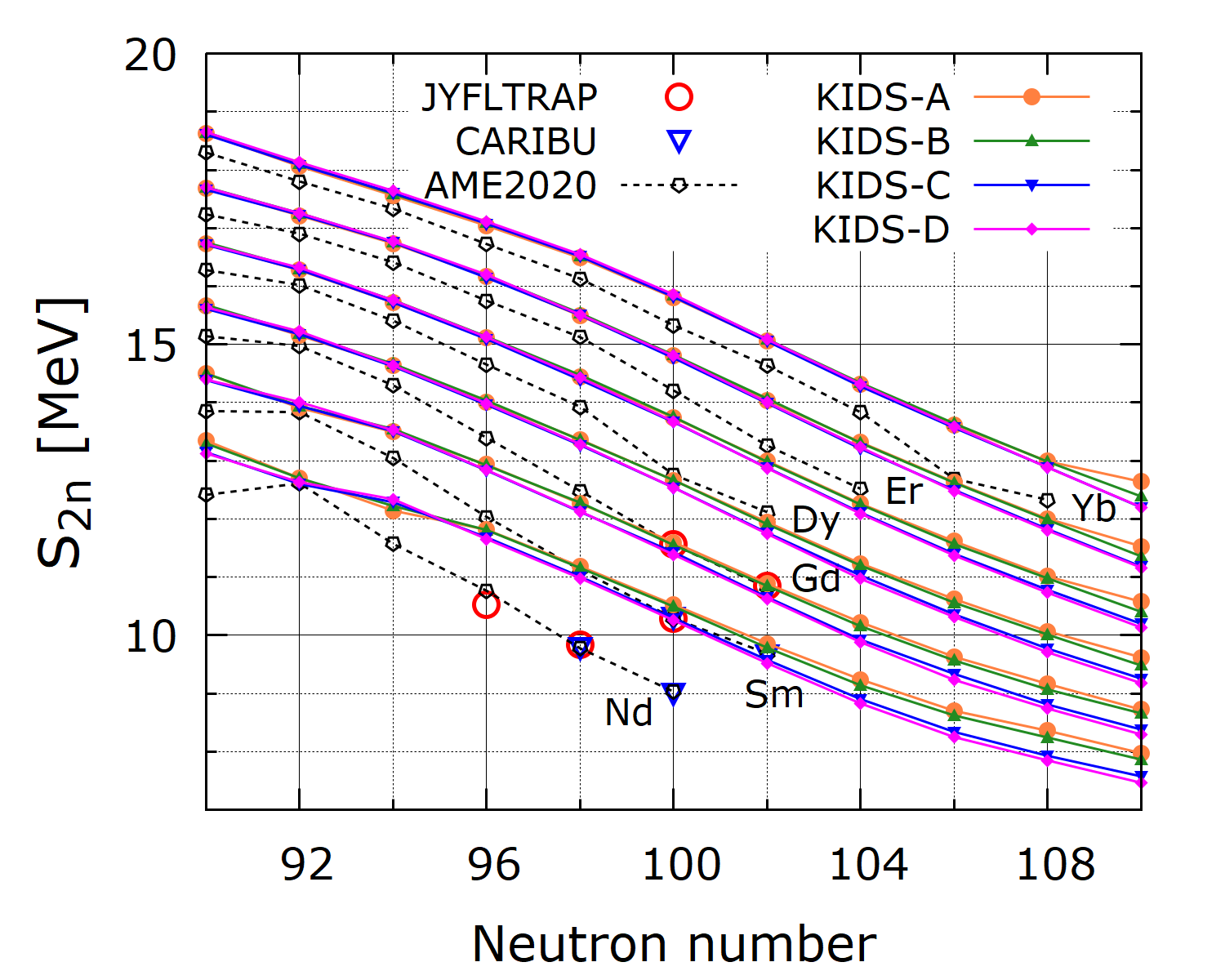} 
\end{tabular}
\caption{Two-neutron separation energy $S_{2n}$
of the rare earth isotopes with $Z=60\textrm{--}70$. 
The experimental data for JYFLTRAP and CARIBU
are obtained from~\cite{JYFLTRAP,CARIBU}, respectively.
\label{fig:S2n}
}
\end{figure}
%%%%%

According to Fig.~\ref{fig:beta}, the rare-earth nuclei with $60 \leq Z \leq 70$ and $90 \leq N \leq 108$ are prolately deformed. 
Thus, the structure change is gradual, and one can expect to see 
a global isospin dependence. 
Then, we show in Fig.~\ref{fig:S2n} 
two-neutron separation energies $S_{2n}$.
The KIDS models overestimate the 
experimental values of $S_{2n}$
and show a smooth decrease as the neutron number increases.
Note 
that the drip line has been confirmed up to the Ne isotopes~\cite{PhysRevLett.123.212501}.
The present KIDS models overestimate the position of the drip line even
in the light region.
Therefore, the neutron drip line of the rare-earth isotopes 
predicted by the KIDS models 
would be located on a more neutron-rich side than where in nature.
In our previous analysis of the Nd isotopes, 
we found that the isotopic dependence of $S_{2n}$ is 
grouped into two: Group 1 and Group 2~\cite{nd2022}; 
the KIDS-A--D models belong to Group 2.
For $Z \sim 70$ isotopes, they show a similar isotopic dependence 
among Group 2. 
As decreasing the proton number,
one can see that KIDS-C and D give a lower $S_{2n}$ value 
than KIDS-A and B do in neutron-rich nuclei.

\begin{figure*}
\begin{center}
\includegraphics[width=0.95\textwidth]{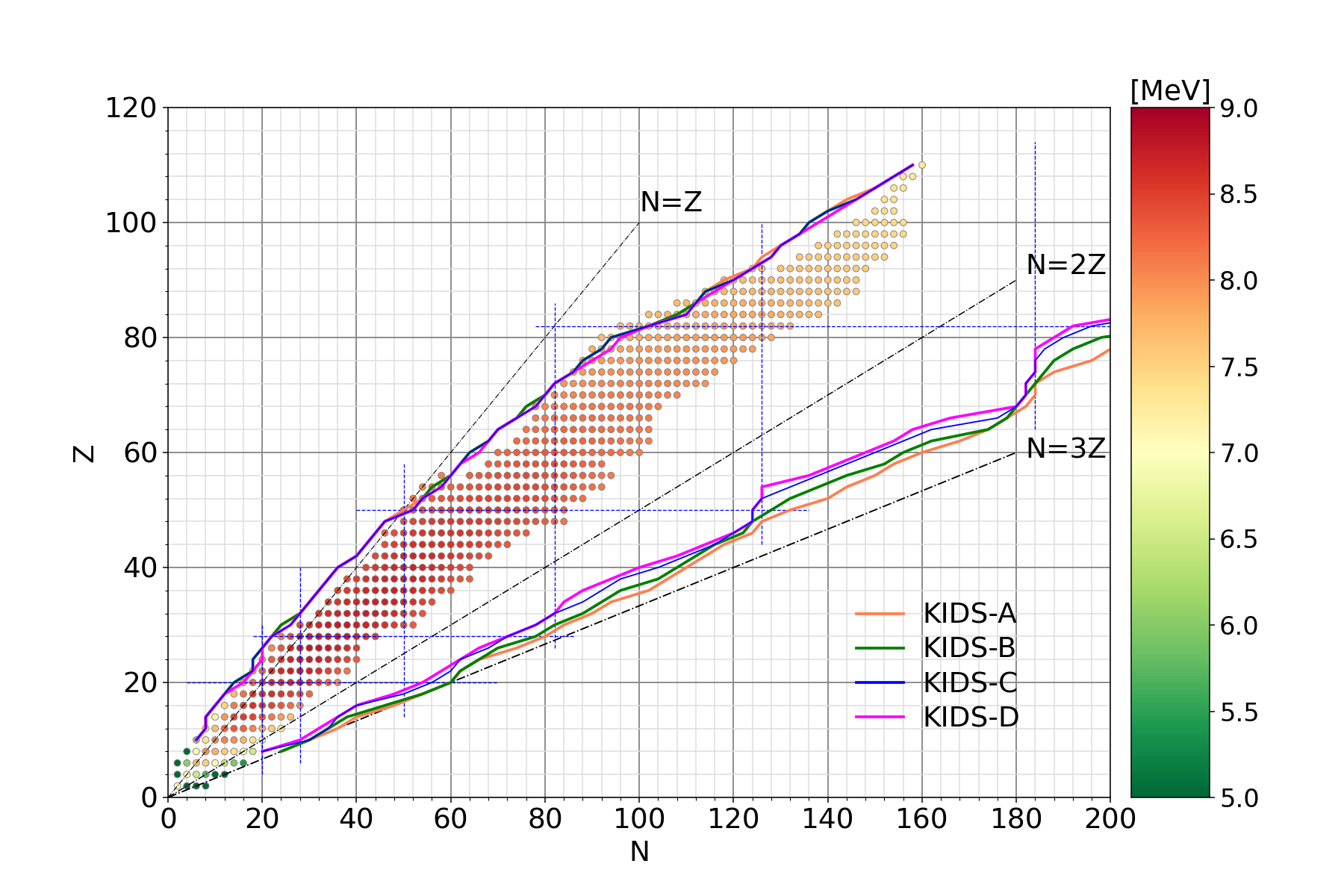}
\end{center}
\caption{
\label{fig:dripline}
Neutron and proton drip lines predicted by the KIDS models.
Filled circles indicate the $E/A$ value taken from AME2020 \cite{ame2020}. 
}
\end{figure*}

Figure~\ref{fig:dripline} 
depicts the predictions for the neutron and proton drip lines.
The drip line is determined by looking at the sign change in the chemical potential.
In spite of the uncertainty of the symmetry energy, proton drip lines are similarly predicted by the models.
When the drip positions differ between the models, the neutron numbers differ only by two at most.
Therefore, models practically predict identical results for the proton drip line.
The reason for the proton drip being insensitive to the symmetry energy is analyzed well with the semi-empirical mass formula in Ref.~\cite{prc2015r}.
The KIDS models also agree well with the experimental data.

The drip line is located in a more neutron-rich region 
with the order KIDS-A $\gtrsim$ KIDS-B $>$ KIDS-C $\gtrsim$ KIDS-D.
The order is the same as the magnitude of the parameter, 
which has been proposed for characterizing the structure of low-mass neutron stars: $\eta_\tau = (-K_\tau L^5)^{1/6}$~\cite{Sotani:2022zvu}.
The $\eta_\tau$ value is 89.8, 80.7, 78.6, and 66.0 MeV for the KIDS-A, B, C, and D, respectively.

Looking into more details, one can see some features specific to each model. As a general trend,
the difference in the position of the neutron drip line between models becomes large with the increase of the proton number.
However, in a certain interval, the difference in the neutron between the models becomes four or less.
Those small uncertainty regions are located at $Z =$ (8, 10, 12, 14), (22, 24), (44, 46, 48), and (68, 70, 72, 74), corresponding to $N\sim28$, 64, 126, and 184.
Small uncertainty regions are also obtained in the other EDFs.
In Ref.~\cite{plb2013}, very small uncertainty happens at $N=126$ and 184,
and the reason is attributed to the spherical shell closure at these magic numbers.
We obtain similar small uncertainties 
around $N=28$ and 64, as well as
$N=126$ and 184.

%%%%%

\section{Summary \label{summary}}
The KIDS-A--D functionals were constructed based on the 
NS observations. 
It is noted that both nuclear matter properties and nuclear data are used
in the decision of the model parameters, and the number of nuclear data used is much smaller
than in the conventional models.
The motivation for this different fitting scheme is to obtain a framework accurate
and applicable to both infinite nuclear matter and finite nuclei.
We have employed these functionals augmented by the mixed-type pairing-density functional to describe the properties of even-even nuclei across the nuclear chart.
We have analyzed the results for the total binding energy, charge radius, quadrupole deformation, 
and the neutron drip line.
The root-mean-square deviation for about 600 nuclei from 
the AME2020 data is as large as $\sim 5$ MeV, 
which is compatible with widely-used nuclear EDFs 
but is larger than recently developed mass models.
We have obtained a similar accuracy for the charge radii to the existing functionals.
The appearance and development of nuclear deformation
in open-shell nuclei are well described 
in spite of no deformed nuclei being considered 
in the construction of the KIDS functional. 
We have found that the location of the neutron drip line is according to the nuclear-matter 
parameter, $\eta_\tau=(-K_\tau L^5)^{1/6}$, 
characterizing the low-mass NS.
The similarity of RMSD values to the widely-used models indicates that a unified description
of nuclear matter and finite nuclei is accessible in the KIDS framework.
The results open a challenge to the KIDS
functional whether it can achieve an accuracy
comparable to the existing globally-fitted EDF models.

\section*{Acknowledgments}
This work was supported by the 
NRF research Grants 
(No. 2018R1A5A1025563 and No. 2023R1A2C1003177), 
JSPS KAKENHI (Grants No. JP19K03824, No. JP19K03872, No. JP19KK0343, and No. JP20K03964), 
and the JSPS/NRF/NSFC A3 Foresight Program ``Nuclear Physics in the 21st Century.''

%\bibliography{ref_KIDS_deformation}

\end{document}